\documentclass[conference]{IEEEtran}

\usepackage{subcaption}
\usepackage[normalem]{ulem}
\usepackage{etex}
\usepackage{mathtools}
\usepackage{tikz}
\usepackage{graphicx}
\usepackage{color}
\usepackage{xcolor}
\usepackage{listings}
\usepackage{flushend}
\usepackage{algorithm2e}
\usepackage{multirow}
\usepackage{multicol}
\usepackage{varwidth}
\usepackage{sidecap}
\usepackage{amsmath}
\usepackage{amssymb}
\usepackage{amsfonts}
\usepackage{enumerate}
\usepackage{xspace}
\usepackage[textsize=tiny, textwidth=1.75cm]{todonotes}
\usetikzlibrary{calc}
\usetikzlibrary{shapes,snakes}
  \usetikzlibrary{positioning,arrows}
\tikzstyle{myarrows2}=[line width=1mm,draw=black,->] 
\tikzstyle{myarrows}=[line width=1mm,draw=blue,-triangle 45, postaction={draw, line width=1.5mm, shorten >=0.3mm, -}]
\tikzstyle{myarrows1}=[line width=1mm,draw=red,  dashed, postaction={draw, dashed, line width=1.1mm}]
\tikzstyle{myarrows3}=[line width=0.5mm, dashed, draw=black,-] 
\newcommand{\commentout}[1]{}

\usepackage{url}
\def\baselinestretch{0.98}

\begin{document}

\title{
Accelerator Codesign as Non-Linear Optimization}

\author{

 \IEEEauthorblockN{Nirmal Prajapati, Sanjay
     Rajopadhye} \IEEEauthorblockA{Colorado State University \\ Fort Collins, CO
 }
 \and
 \IEEEauthorblockN{Hristo Djidjev, Nandkishore Santhi}
 \IEEEauthorblockA{Los Alamos National Laboratory\\ Los Alamos, NM, USA}
 \and
 \IEEEauthorblockN{Tobias Grosser}
 \IEEEauthorblockA{ETH \\ Zurich}
 \and
 \IEEEauthorblockN{Rumen Andonov}
 \IEEEauthorblockA{IRISA \\ University of Rennes}
}

\maketitle

\begin{abstract}

  We propose an optimization approach for determining both hardware and
software parameters for the efficient implementation of a (family of)
applications called \emph{dense stencil computations} on programmable GPGPUs.
We first introduce a simple,
  analytical model for the silicon area usage of accelerator architectures and
a workload characterization of stencil computations.  We combine this
characterization with a
  parametric execution time model and formulate a mathematical
  optimization problem.  That problem seeks to maximize a common objective
  function of \emph{all the hardware and software parameters}.  The solution
  to this problem therefore ``solves'' the codesign problem: simultaneously
  choosing software-hardware parameters to optimize total performance.

  We validate this approach by proposing architectural variants of the NVIDIA
  Maxwell GTX-980 (respectively, Titan~X) specifically tuned to a
  predetermined workload of four common 2D stencils (Heat, Jacobi, Laplacian,
  and Gradient) and two 3D ones (Heat and Laplacian).  Our model predicts that performance would potentially
  improve by 28\% (respectively, 33\%) with simple tweaks to the hardware
  parameters such as adapting coarse and fine-grained parallelism by changing
  the number of streaming multiprocessors and the number of compute cores each
contains.
  We propose a set of Pareto-optimal design points to exploit the trade-off
  between performance and silicon area and show that by additionally
eliminating GPU caches, we can get a further 2-fold improvement.

\end{abstract}

\IEEEpeerreviewmaketitle

\section{Introduction}

Software-hardware codesign is one of the proposed enabling technologies for
exascale computing and beyond~\cite{b1282de6ab8a4f1a8a962cb40310bc7f}.
Currently, hardware and software design are done largely separately.  Hardware
manufacturers design and produce a high-performance computing (HPC) system
with great computing potential and deliver it to customers, who then try to
adapt their application codes to run on the new system.  But because of a
typically occurring mismatch between hardware and software structure and
parameters, such codes are often only able to run at a small fraction of the
total performance the new hardware can reach.  Hence, optimizing both the
hardware and software parameters \textit{simultaneously} during hardware
design is considered as a promising way to achieve better hardware usage
efficiency and thereby enabling leadership-class HPC availability at a more
manageable cost and energy efficiency.

The design of HPC systems and supercomputers is by no means the only scenario
where such optimization problems occur.  The execution platforms of typical
consumer devices like smart phones and tablets consist of very heterogeneous
Multi-Processor Systems-on-Chip (MPSoCs) and the design challenges for them are
similar.

Despite the appeal of an approach to \emph{simultaneously} optimize for
software and hardware, its implementation represents a formidable challenge
because of the huge search space.  Previous approaches
\cite{codesign-white-paper, opt_principles,
  Mniszewski:2015:TDE:2764453.2699715}, pick a hardware model ${\cal H}$ from
the hardware design space, a software model ${\cal S}$ from the software
design space, map ${\cal S}$ onto ${\cal H}$, estimate the performance of
the mapping, and iterate until a desirable quality is achieved. But not only
each of the software and hardware design spaces can be huge, each iteration
takes a long time since finding a good mapping of ${\cal S}$ onto ${\cal H}$
and estimating the performance of the resulting implementation are themselves
challenging computational problems.

In this paper, we propose a new approach for the software-hardware codesign
problem that avoids these pitfalls by considerably shrinking the design space
and making its exploration possible by formulating the optimization problem in
a way that allows the use of existing powerful optimization solvers.  We apply
the methodology to programmable accelerators: Graphics Processing Units
(GPUs), and for stencil codes.  The key elements of our approach are to
exploit multiple forms of \emph{domain-specificity}.
Our main contributions are:
\begin{itemize}
\item We propose a new approach (see Section~\ref{sec:approach}) to
  software-hardware codesign that it is computationally feasible and provides
  interesting insights.
\item We develop a simple, analytical model for the silicon area
  (Section~\ref{sec:area}) of programmable accelerator architectures, and
  calibrate it using the NVIDIA Maxwell class GPUs.
\item We combine this area model with a workload characterization of stencil
  codes, and our previously proposed execution time
  model~\cite{prajapati2017simple} to formulate a mathematical optimization
  problem that maximizes a common objective function of the hardware and
  software parameters (see Section~\ref{sec:codesign}).
\item Our analysis (Section~\ref{sec:discussion}) provides interesting
  insights.  We produce a set of Pareto optimal designs that represent the
  optimal combination of hardware and compiler parameters.  They allow for up to
  $33\%$ improvement in performance as measured in
  GFLOPs/sec.
\end{itemize}




\section{Approach}
\label{sec:approach}

The key element of our approach is exploiting \emph{domain specificity}.  We
do this in three ways.  First, we tackle a specific (family of) computations
that are nevertheless very important in many embedded systems.  This class of
computations, called \emph{dense stencils}, includes the compute intensive
parts of many image processing kernels, simulation of physical systems
relevant to realistic visualization, as well as the solution of partial
differential equations (PDEs) that arise in many cyber-physical systems such
as automobile control and avionics.

Second, we target GPU-like \emph{vector-parallel programmable accelerators}.
Such components are now becoming de-facto standard in most embedded platforms
and MPSoCs since they provide lightweight parallelism and energy/power
efficiency.  We further argue that they will become ubiquitous for the
following reasons.  Any device on the market today that has a screen
(essentially, \emph{any device}, period) has to render images.  GPUs are
natural platforms for this processing (for speed and efficiency).  So all
systems will have an accelerator, by default.  If the system now needs any
additional dense stencil computations, the natural target for performing it in
the most speed/power/energy efficient manner is on the accelerator.

The third element of domain specificity is that we exploit a formalism called
the \emph{polyhedral model} as the tool to map dense stencil computations to
GPU accelerators.  Developed over the past thirty years~\cite{sanjay-fst-tcs,
  quinton-jvsp89, feautrier91, feautrier92a, feautrier92b}, it has matured into
a powerful technology, now incorporated into \texttt{gcc}, \texttt{llvm} and
in commercial compilers [Rstream, IBM].  Tools targeting GPUs are also
available~\cite{grosser-etal-GPUhextile-CGO2014, chen_etal08_ChillTR}.  We
recently showed~\cite{prajapati2017simple} that these elements of the domain
specificity can be combined to develop a simple analytical model for the
execution time of tiled stencil codes on GPUs, and that this model can be used
to solve for optimal tile size selection.

Thus, we formulate the domain specific optimization problem as:
\emph{simultaneously optimize compilation and hardware/architectural
  parameters for compiling stencil computations to GPU-like vector-parallel
  accelerators}.

To explain our overall approach, first recall how tile size selection is
formulated as an optimization problem~\cite{prajapati2017simple}.  A given
class of programs, with some \emph{problem parameters},
$\overrightarrow{p} \in {\cal P}$ (e.g., size of the iteration space, number
of array variables) is compiled to an accelerator with some \emph{hardware
  parameters} $\overrightarrow{h} \in {\cal H}$ (e.g., number of cores, number
of vector units, cache size, etc.)  The compiler itself may use some well
defined optimization strategies (e.g., tiling, skewing, parallelization,
etc.), and may have some \emph{software parameters}, notably tile sizes
($\overrightarrow{s}$).

We develop an analytical function,
$\mathcal {T}(\overrightarrow{p}, \overrightarrow{h}, \overrightarrow{s})$
that predicts the execution time of the target program as a function of these
parameters.  The values of the parameters must satisfy a set of
\emph{feasibility constraints} (e.g., tile sizes must be such that the data
footprint of a tile must fit in the available scratchpad memory) denoted by
$\mathcal{F}(\overrightarrow{s}, \overrightarrow{h}, \overrightarrow{s})$.
Then, optimal tile size selection is simply solving the following problem:
\begin{align}
  \underset{\overrightarrow{s}} {\text{minimize~~}}
  && \mathcal{T}(\overrightarrow{p}, \overrightarrow{h}, \overrightarrow{s})
     \label{eq:tss} \\
  \text{subject to:}
  && \overrightarrow{s} \in \mathcal{F}(\overrightarrow{p},
    \overrightarrow{h}, \overrightarrow{s})\nonumber
\end{align}

Here, the unknown variables are $\overrightarrow{s}$.  The hardware and
program parameters are considered as constants---compiling a different program
and/or compiling to a different target machine entails solving a new
optimization problem.  Extending this to codesign requires the following
steps.
\begin{itemize}
\item \emph{Workload characterization} We first pick a set $\mathcal{W}$ of
  representative program instances.  For each one, we use profiling to pick
  the probability/frequency with which it occurs in the workload.  Each
  program may be executed with a range of program parameters, and we also have
  a frequency with which of these appear in the workload.  So,
  $\overrightarrow{p_{i,j}}$ are the program parameters of the $i$-th instance
  of the $j$-th benchmark, and the set
  $\{\overrightarrow{p_{i,j}}~|~ i \in \mathrm{Instances}, j \in
  \mathrm{Benchmarks}\}$ is $\mathcal{P}$, the range of program parameters.
\item \emph{Codesign Optimization} Now we formulate a simple extension of
  (\ref{eq:tss}).  Instead of leaving the hardware parameters as constants, we
  allow them to be unknown variables of the optimization problem, which now
  becomes
\begin{align}
  \underset{\langle \overrightarrow{s} \overrightarrow{h}\rangle}
  {\text{minimize~~}}
  && \mathcal{T}(\overrightarrow{p}, \overrightarrow{h}, \overrightarrow{s})
     \label{eq:tss} \\
  \text{subject to:}
  && \overrightarrow{s} \in \mathcal{F}(\overrightarrow{p},
    \overrightarrow{h}, \overrightarrow{s})\nonumber \\
  && \overrightarrow{h} \in \mathcal{F}_h(\overrightarrow{h})\nonumber
\end{align}
\item \emph{Area Model} Since $\overrightarrow{h}$ is now an unknown, we must
  formulate its feasible space, $\mathcal{F}_h (\overrightarrow{h})$
  precisely.  For this, we develop an analytical model of the area of the
  accelerator.  We assume that we are given an area budget within which the
  accelerator must fit, and solve the optimization problems over the resulting
  feasible space.
\end{itemize}

This seems (deceptively) simple, but the devil is in the detail.  The
resulting optimization problem has many hundreds of variables, and has
non-convex constraints and objective functions, making it computationally
intractable.  The details of how we solve it are explained in
Section~\ref{sec:codesign}.


\section{Area Model}
\label{sec:area}

We now develop an analytic model for the total silicon area of a GPU
accelerator.  We faced some difficulties in deriving an acceptable analytical
model, as silicon data had to be reverse engineered from extremely limited
public domain resources.  As a general observation, within each GPU family,
there is little diversity in the parameter configurations.  For the Maxwell
family of GPUs, the GTX980 and Titan~X chips were chosen as two sufficiently
distinct points to calibrate our analytical models.  The calibration itself
was performed by evaluating die photomicrographs, publicly available
information about the nVidia GTX-980 (Maxwell series) GPU, and other generally
accepted memory architecture models.  The model validation was done by
comparing the predictions with known data on the Maxwell series Titan X GPU.
We found the model prediction to be accurate to within, 2\%, though this
number is not significant.\footnote{Although a many configurations of any
  family of GPUs are spaced out, they come from binning only a small number of
  distinct dies.  We ended up calibrating our model on one die and validating
  it on only another one.}

\subsection{Analytical Model for GPU Area}

The area model is a function of the main parameters that we want to select
optimally in the codesign problem: the number of SMs, $n_\mathrm{SM}$, the
number of vector units or cores in each SM, $n_\mathrm{V}$, and the respective
sizes of the register file and shared memory size of the SM, $R_\mathrm{SM}$,
and $M_\mathrm{SM}$.  Table~\ref{tab:areapars} shows a summary of the various
parameters used in developing the area model.

\begin{table}
  \centering
  \scriptsize
  \begin{tabular}{|l|l|}
    \hline
    Name & Description \\
    \hline\hline
    $\alpha_R$ & overhead area per kB of register-memory per vector-unit \\
    \hline
    $\alpha_M$ & overhead area per kB of shared-memory per SM \\
    \hline
    $\alpha_{L1}$ & L1 cache overhead area per SM-pair \\
    \hline
    $\alpha_{L2}$ & L2 cache overhead area \\
    \hline
    $\alpha_{oh}$ & common overhead area (I/O, global routing etc) per SM \\
    \hline
    $\beta_R$ & area per register-file-bank per kB per vector-unit \\
    \hline
    $\beta_M$ & area per shared-memory-bank per kB per SM \\
    \hline
    $\beta_{L1}$ & L1 cache area per kB per SM-pair \\
    \hline
    $\beta_{L2}$ & L2 cache area per kB \\
    \hline
    $\beta_{VU}$ & core-logic area within a vector-unit \\
    \hline\hline
    $n_\mathrm{SM}$ & total number of SM on the GPU chip \\
    \hline
    $n_\mathrm{V}$ & number of vector-units per SM \\
    \hline
    $R_\mathrm{VU}$ & kB of register files per vector-unit \\
    \hline
    $M_\mathrm{SM}$ & kB of shared memory per SM \\
    \hline
    $L1_\mathrm{SMpair}$ & kB of L1 cache per SM-pair \\
    \hline
    $L2_\mathrm{SM}$ & kB of L2 cache \\
    \hline\hline
    $\mathcal{A}_\mathrm{tot}$ & total GPU chip die area \\
    \hline
    $\mathcal{A}_\mathrm{SM}$ & total shared-memory die area \\
    \hline
    $\mathcal{A}_\mathrm{cache}$ & total cache die area \\
    \hline
    $\mathcal{A}_\mathrm{oh}$ & total on-chip overhead die area \\
    \hline
    $\mathcal{A}_\mathrm{LSU}$ & total load-store unit die area \\
    \hline
    $\mathcal{A}_\mathrm{SFU}$ & total special-function unit die area \\
    \hline
    $\mathcal{A}_\mathrm{FDU}$ & total fetch-decode unit die area \\
    \hline
    $\mathcal{A}_\mathrm{Icache}$ & total instruction-cache die area \\
    \hline\hline
    $\mathcal{A}_\mathrm{LSUperSM}$ & load-store unit die area per SM \\
    \hline
    $\mathcal{A}_\mathrm{LSUperV}$ & load-store unit die area per vector-unit \\
    \hline
    $\mathcal{A}_\mathrm{MperSM}$ & memory die area per SM \\
    \hline
    $\mathcal{A}_\mathrm{SFUperV}$ & special-function unit die area per vector-unit \\
    \hline\hline
  \end{tabular}
    \vspace{0.5em}
    \caption{Area model parameters.  The top two groups are elementary
      parameters, and the third one is composite (some function of the
      elementary parameters).  Of the elementary ones, the second one are
      treated as variables in our optimization formulation.}
    \label{tab:areapars}
\end{table}

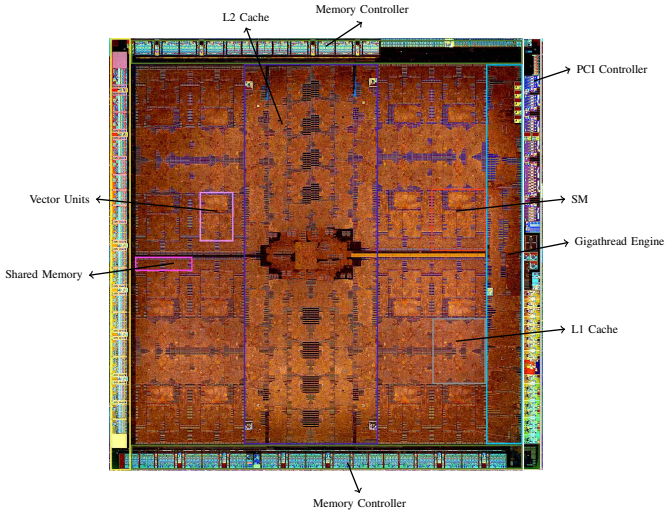
\begin{figure}[tph]
  \begin{subfigure}[b]{0.5\textwidth}
    \centering \resizebox{\linewidth}{!}{
      \begin{tikzpicture}
        \node[anchor=south west,inner sep=0] (image) at (0,0,0)
        {\includegraphics[width=\textwidth]{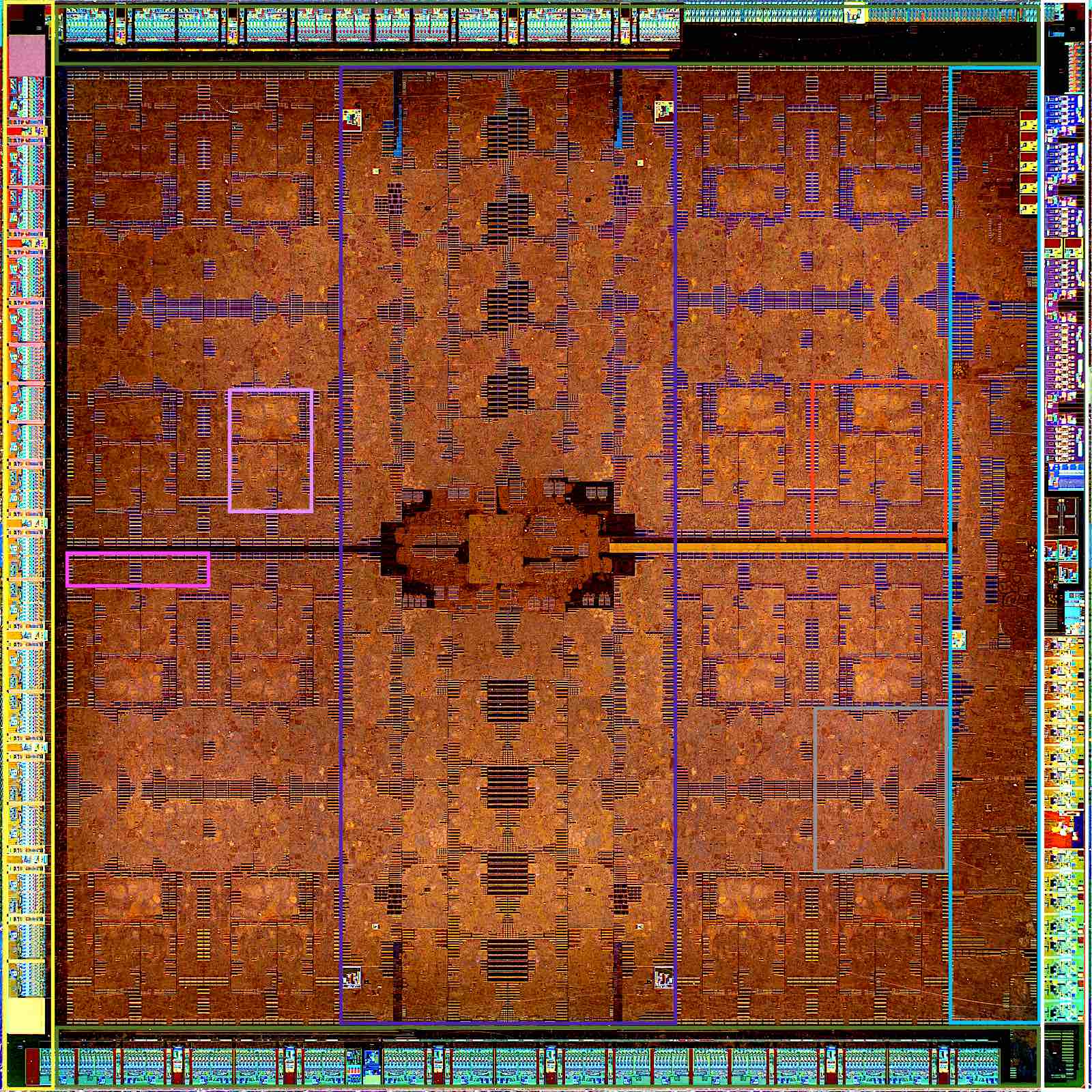}};
        \begin{scope}[x={(image.south east)},y={(image.north west)}]
          \draw[thick,solid,->] (0.55,0.02) --
          +(0.1in,-0.25in)node[anchor=north] {\scriptsize Memory Controller};
          \draw[thick,solid,->] (0.5,0.98) -- +(0.3in,0.2in)node[anchor=south]
          {\scriptsize Memory Controller}; \draw[thick,solid,->] (0.98,0.9) --
          +(0.3in,0.1in)node[anchor=west] {\scriptsize PCI Controller};
          \draw[thick,solid,->] (0.92,0.5) -- +(0.5in,0.1in)node[anchor=west]
          {\scriptsize Gigathread Engine}; \draw[thick,solid,->] (0.8,0.3) --
          +(0.9in,0.1in)node[anchor=west] {\scriptsize L1 Cache};
          \draw[thick,solid,->] (0.4,0.8) -- +(-0.3in,0.8in)node[anchor=south]
          {\scriptsize L2 Cache}; \draw[thick,solid,->] (0.15,0.48) --
          +(-0.7in,-0.1in)node[anchor=east] {\scriptsize Shared Memory};
          \draw[thick,solid,->] (0.25,0.6) -- +(-1.0in,0.1in)node[anchor=east]
          {\scriptsize Vector Units}; \draw[thick,solid,->] (0.8,0.6) --
          +(0.9in,0.1in)node[anchor=west] {\scriptsize SM};
        \end{scope}
      \end{tikzpicture}
    }
  \end{subfigure}
  \protect\caption{Best effort annotation of functional blocks in a GTX980
    die photograph~\cite{Fritz} based on current understanding of the nVidia Maxwell architecture and
    chip layout. 
    }
  \label{fig:GM204:annotated}
\end{figure}

\vspace*{-0.5ex}
\begin{equation}
    \label{eq:area1-acc}
    \mathcal{A}_\mathrm{tot} () = n_\mathrm{SM} \mathcal{A}_\mathrm{SM} + \mathcal{A}_\mathrm{cache} + \mathcal{A}_\mathrm{oh}
\end{equation}
where $\mathcal{A}_\mathrm{SM}$ is the cost of each SM,
$\mathcal{A}_\mathrm{cache}$ is the total area of the on-chip L1 and L2
caches, and $\mathcal{A}_\mathrm{oh}$ is the overall overhead area cost.  The
term $\mathcal{A}_\mathrm{oh}$ accounts for several components which are of
only peripheral interest to us, such as metal-layer routing overheads, I/O
pads, buffers, and clock-signal global distribution trees, gigathread
scheduler, PCI express interface, raster engine, and memory controller.  Based
on our current understanding of the nVidia architecture, a best effort
annotation of various functional blocks is shown overlaid on a die
photomicrograph of GTX980 chip in Figure~\ref{fig:GM204:annotated}.  The term
$\mathcal{A}_\mathrm{SM}$ denotes the area of a single SM, and we develop it
next.  The following micro-architectural elements comprise an SM:

\begin{itemize}\itemsep 0mm
\item Individual vector units: The area cost of this is
  $n_\mathrm{V} \beta_{VU}$.
\item Load-store unit (LSUs): Every core may issue independent memory
  requests, to either the shared memory or to global memory.  Therefore the
  LSU has a component that is replicated per core, and also a shared component
  that (eventually) collects together all requests from a single warp and
  interfaces to the (global) memory hierarchy.  Therefore
  $\mathcal{A}_\mathrm{LSU} = n_\mathrm{V} \mathcal{A}_\mathrm{LSUperV} +
  \mathcal{A}_\mathrm{LSUperSM}$.
\item Special function units (SFUs): These are dedicated functional units that
  are used for common graphics functions, transcendentals, etc.  Their number
  is not always exactly equal to $n_\mathrm{V}$, (e.g., in the nVidia GTX-980,
  there is one SFU every 8~cores) but we model it as a linear function of
  $n_\mathrm{V}$, i.e.,
  $\mathcal{A}_\mathrm{SFU} = n_\mathrm{V} \mathcal{A}_\mathrm{SFUperV}$.
\item Instruction fetch-decode unit (FDU): Like the I-cache, there is a single
  FDU per SM, so its cost is simply a constant $\mathcal{A}_\mathrm{FDU}$
\item Various specialized ``memories:'' the shared memory used as scratchpad,
  the register file, the instruction cache, the texture and other special
  caches.  We call this term $\mathcal{A}_\mathrm{MperSM}$ and develop it
  separately.
\end{itemize}

  Of these, the memories other than register files and the FDU are shared by the
entire SM, while the register files, SFU and LSU are accounted for on a per vector unit basis.
Hence,
\vspace*{-0.5ex}

\begin{eqnarray}
    \label{eq:area1-SM}
    \mathcal{A}_\mathrm{SM} ()  =  (\mathcal{A}_\mathrm{FDU} + \mathcal{A}_\mathrm{Icache} + \mathcal{A}_\mathrm{LSUperSM}) + \mathcal{A}_\mathrm{MperSM} \nonumber \\ 
     + n_\mathrm{V} (\mathcal{A}_\mathrm{SFUperV} + \mathcal{A}_\mathrm{core} + \mathcal{A}_\mathrm{LSUperV}) \nonumber \\ 
     = \alpha' + \mathcal{A}_\mathrm{MperSM} + \beta n_\mathrm{V}
\end{eqnarray}
where
$\alpha' = \mathcal{A}_\mathrm{FDU} + \mathcal{A}_\mathrm{Icache} +
\mathcal{A}_\mathrm{LSUperSM}$, and $\mathcal{A}_\mathrm{MperSM}$ depends on
the capacity of each of the individual memories.  Of these, we are interested
in modeling the area costs for the register file, shared (scratchpad) memory,
L1 cache and L2 cache; so the others are treated as constants absorbed in
$\alpha'$.  L1 cache is shared by a pair of SM units, while the L2 cache is
shared by all the SM units on the chip.  The register files are organized so
that a few registers (512 registers each 32 bits wide in GTX980) are
exclusively accessible by each vector-unit.  The shared memory is accessible
by all vector-units in an SM.  Assuming independent linear cost models for
each of these four memory types, therefore,
$\mathcal{A}_\mathrm{MperSM} () = n_\mathrm{V} \left(\beta_\mathrm{R}
  R_\mathrm{VU} + \alpha_\mathrm{R} \right) + \left(\beta_\mathrm{M}
  M_\mathrm{SM} + \alpha_\mathrm{M} \right) + \tfrac{n_\mathrm{SM}}{2}
\left(\beta_\mathrm{L1} L1_\mathrm{SMpair} + \alpha_\mathrm{L1} \right) +
\left(\beta_\mathrm{L2} L2_\mathrm{SM} + \alpha_\mathrm{L2}
\right)$.  Here, we have subscribed to a design philosophy where the size of
the L2 cache is not proportional to the number of SM units, it is a
constant.  This choice
seems to be the norm, in that the GTX980 chip with $16$ SMs has an L2 cache of
$2$MB, whereas the Titan X with 24 SMs has an L2 cache of $3$MB.  Another
design choice we make is the following -- the common area overhead
$\mathcal{A}_{oh}$, comprising the I/O pads, buffers, routing, gigathread
engine, PCI and memory controllers, is also assumed to be proportional to the
number of SMs\footnote{This design choice is by no means the only possible one
  -- another possibility is to also have an additional constant term, though
  one would need more than two GPUs in the Maxwell family with different die
  areas to calibrate such a model.}, that is,
$\mathcal{A}_{oh} = n_\mathrm{SM} \alpha_{oh}$.  Substituting this result in
the above equations, collecting all common overhead area contributions
including $\alpha'$ into $\alpha_{oh}$, and further simplifying, we get,
\begin{eqnarray}
    \mathcal{A}_\mathrm{tot} ()  \,=\, 
      n_\mathrm{SM} n_\mathrm{V} \beta_{VU}
    + n_\mathrm{SM} n_\mathrm{V} \left(\beta_\mathrm{R} R_\mathrm{VU} + \alpha_\mathrm{R} \right)
      \nonumber \\  
    + n_\mathrm{SM} \left(\beta_\mathrm{M} M_\mathrm{SM} + \alpha_\mathrm{M} \right)
    + \tfrac{n_\mathrm{SM}}{2} \left(\beta_\mathrm{L1} L1_\mathrm{SMpair} + \alpha_\mathrm{L1} \right) 
      \nonumber \\
    + \left(\beta_\mathrm{L2} L2_\mathrm{SM} + \alpha_\mathrm{L2} \right)
    + n_\mathrm{SM} \alpha_{oh}
\end{eqnarray}

\subsection{Calibrating the Model}
Since the overall area model involves non-linear terms, we used an incremental
fitting approach to calibrate our model.  We first developed linear regression
models for the area contribution due to the memory elements.  We then
incorporated the memory element area model into a measurement based linear
model for the area due to the SM core vector units.

The Maxwell family was fabricated using the TSMC $28$nm process.  For this
process, the typical SRAM bit cell has an area of only $0.127$ to $0.155$
$\mu m^2$ \cite{tsmc-28nm-sram}.  The total area used by memory banks, be it
the register files, shared memory, or the caches, involve the much more
substantial additional area used for routing, addressing, read, write and
bit-sense amplification logic.  The register file, shared memory, and cache
area models should take this overhead into account, by accounting for the bit
width, number of addresses, and number of read and write ports.  We used the
open source Cacti 6.5 RAM and cache estimation tool from HP labs \cite{cacti}
to optimize and calibrate all our memory models:

\begin{itemize}\itemsep 0mm
\item Register files: We modeled the register file at the vector-unit level
  with a data bus of $32$ bits.  For example, the GTX980 chip has 512
  such registers per vector-unit. For configuring the register file in
  Cacti, we assumed a `ram' model which is direct-mapped, with $2$
  exclusive single ended read ports and $1$ exclusive write port per
  vector-unit.  Our Cacti design objective was to aggressively minimize area,
  as register files are placed physically close to the logic core where space
  is at an extra premium.

\item Shared memory units: The shared memory bank is organized per SM.  The
  GTX980 chip has $96$kB of shared memory per SM.  The user is allowed to
  optionally dynamically configure a part of the shared memory in an SM as
  cache.  We assumed, for the purpose of Cacti tool configuration, that it is
  a direct mapped `ram' type memory with $32$ bit wide data bus, on each of
  its $8$ read-write ports per SM.  Our design objective was to minimize area,
  with a secondary objective of minimizing propagation delays.

\item L1 cache: The L1 cache is shared by two adjacent SM units.  The GTX980
  has $48$kB of L1 cache per SM.  Our Cacti model for L1 cache is a `cache'
  type memory with a line-size of $128$ bytes, and a data width of $32$ bits.
  We also chose a conservative setting of full associativity for this cache
  level.  There are $8$ exclusive read ports for downstream access from the
  logic cores, and $8$ exclusive write ports for upstream access from the L2
  cache.  The design was tailored for speed and the optimization objective was
  primarily propagation delay and secondarily power dissipation.

\item L2 cache: For the nVidia Maxwell architecture, the L2 cache is shared by
  all the SM units on the chip.  For this GPU family, the L2 cache is on
  average $128$kB per SM.  Our Cacti configuration assumes a `cache' type
  memory with a line-size of $128$ bytes, and a data width of $256$ bits on
  each of its $8$ exclusive read ports feeding into every L1 cache on the chip
  downstream.  There is also a exclusive read-write port for upstream
  interface with the memory controllers.  The design objective was a weighted
  mix of delay and area.

\end{itemize}

Using these Cacti models, we obtained area estimates for per vector-unit
register file banks of $512, 1024, 2048, 4096$, and $8192$ bytes each.  We
then derived a linear fit model for the area given the bank size.  The
coefficients of this model were
$\beta_R  = 0.004305$ and $\alpha_R  = 0.001947$.

Similarly, we obtained area estimates for per SM shared memory banks of
$24, 48, 96, 192$, and $384$ kilo-bytes size each, and derived a linear fit
model.  The coefficients of this linear model were
$\beta_M = 0.01565$ and $\alpha_M = 0.09281$.

\begin{figure}[tph]
  \centering
  \includegraphics[width=2.8in]{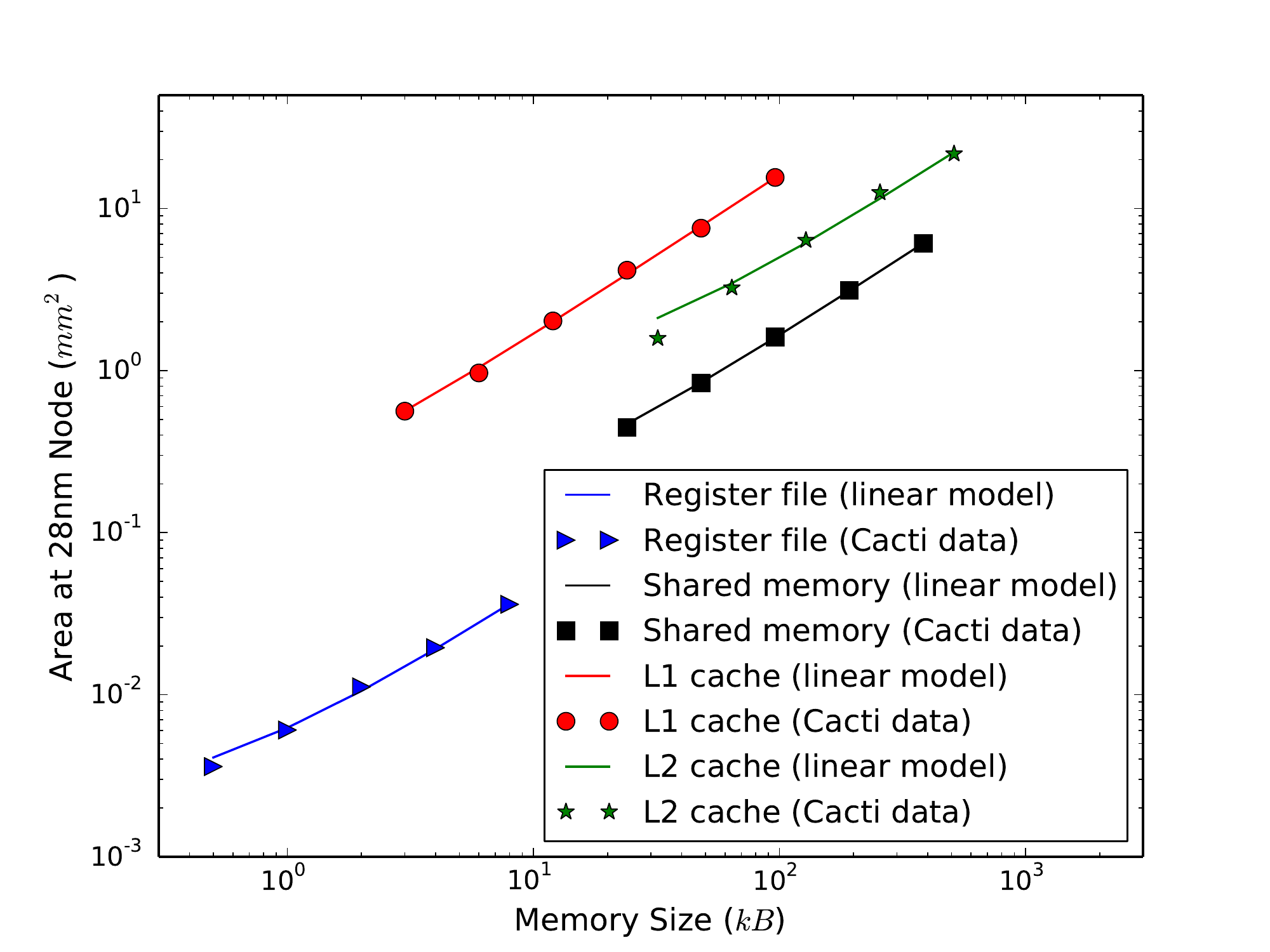}
  \protect\caption{Linear regression models for various memory types on a
    Maxwell series GPU.  The Cacti predicted area used to obtain the linear
    model are also shown. \label{graph:cactimodel}}
\end{figure}

The L1 and L2 cache models were similarly calibrated by performing
optimizations using our Cacti models as described earlier.  The L1 linear fit
model was obtained with per SM-pair sizes of $3, 6, 12, 24, 48$, and $96$
kilo-bytes each.  The model parameters obtained were
$\beta_{L1} = 0.1604$ and $\alpha_{L1} = 0.08204$.  The L2 linear fit model
was obtained with per SM sizes of $32, 64, 128, 256$, and $512$ kilo-bytes
each.  The parameters obtained were
$\beta_{L2} = 0.04197$ and $\alpha_{L2} = 0.7685$.  The linear regression
models obtained as above are shown in Figure~\ref{graph:cactimodel}.

nVidia publishes obfuscated {\em die-shots} of their GPU chips, though these
photographs are sometimes not to scale.  The community of GPU enthusiasts and
researchers have also attempted to decap packaged GPU chips and photograph the
silicon dies independently.  For example, Fritzchens~Fritz~\cite{Fritz} has a
vast collection of nVidia die photographs which we have used in association
with the obfuscated die shots published by nVidia.  The nVidia official
``die-shots'' tend to show demarcated {\em logical} functional blocks - for
SMs, this includes not only the SM core logic regions, but also the associated
control, and thread-scheduler logic areas too.  We have factored these aspects
into account while making our area measurements.  The area measurements were
initially made in terms of pixels, and later normalized to $\text{mm}^2$ using
the total die area from the GPU's datasheet.  As a verification of the memory
area model calibration, we first measured the area of the following memory
blocks on the die photograph -- L2: $105\,\text{mm}^2$, L1:
$7.34\,\text{mm}^2$, and shared memory: $1.27\,\text{mm}^2$.  These measured
areas matched quite well with the linear model predictions -- L2:
$98.25\,\text{mm}^2$, L1: $7.78\,\text{mm}^2$, and shared memory:
$1.59\,\text{mm}^2$.  Furthermore, from the die photomicrographs for GTX980
\cite{Fritz,gtx980die}, area per vector-unit logic core was measured to be
$\beta_{VU} = 0.04282\,\text{mm}^2$, excluding the register-file area.
Similarly, measurements on GTX980 die photograph gave an area estimate of
$102.65\,\text{mm}^2$ for the overhead region containing the I/O pads,
buffers, memory controllers, gigathread and raster engines and PCI controller,
which gives an $\alpha_{oh} = 6.4156\,\text{mm}^2$ per SM.  The total
published die area for GTX980 is also known to be
$\mathcal{A}_\mathrm{tot} = 398\,\text{mm}^2$.  For the Maxwell family, our
overall area model is therefore given by:
\begin{eqnarray}
    &\mathcal{A}_\mathrm{tot} =
      0.0447 n_\mathrm{SM} n_\mathrm{V}
    + 0.0043 R_\mathrm{VU} n_\mathrm{SM} n_\mathrm{V}
    \nonumber\\
    & + 0.015 M_\mathrm{SM} n_\mathrm{SM}
	+ 0.08 L1_\mathrm{SMpair} n_\mathrm{SM}
	\nonumber\\
    & + 0.041 L2_\mathrm{kB}
    + 7.317 n_\mathrm{SM} \label{eq:area-final}
\end{eqnarray}

\subsection{Validating the Model}
In order to validate our area model, we applied it to another member of the
Maxwell family, the Titan X for which the total die area is known.  Our model
predicts a total area of $589.2\,\text{mm}^2$ which is within an error of
$1.96\%$ of the published die area of $601\,\text{mm}^2$.

Our area model above is only valid for the functional blocks as implemented
for the specific TSMC $28$-nm process used to fabricate the nVidia Maxwell
chips.  The newer Pascal family of nVidia GPUs are fabricated on a $16$-nm
technology node.  If each of the constituent functional block's chip level
layouts were to remain same across technology nodes, and were only optically
reduced down by a certain shrinkage factor, then the area model can be
similarly rescaled and reused.  However, the shift from $28$-nm to $16$-nm
involves a non-trivial move from planar CMOS to 3D-FINFET technology, and
consequently a simple area rescaling would clearly not be sufficient to
predict the area parameters of Pascal family chips.  However the overall
methodology still remains applicable to any technology node or device family.


%
%


\section{Optimizing software and hardware parameters}
\label{sec:codesign}
\newcommand{\sz}{\mathrm{Sz}}
\newcommand{\Sz}{\mathrm{SZ}}
\newcommand{\fr}{\mathrm{fr}}
\newcommand{\Ts}{\mathit{SP}} 
\newcommand{\Cd}{\mathrm{Cd}}

We now show how to use the analytical chip area model
and our previous execution time model~\cite{prajapati2017simple} to formulate
and solve an optimization problem for simultaneously finding optimal hardware
and software parameters.

\subsection{Problem formulation}
The parameters we are going to optimize are the elementary
software (ES) parameters $t_\mathrm{S_1}$, $t_\mathrm{S_2}$, and
$t_\mathrm{T}$ and the elementary hardware (EH) parameters $n_\mathrm{SM}$,
$n_\mathrm{V}$, and $M_\mathrm{SM}$.  For each combination of EH parameters,
the corresponding running time in the objective function will be defined as
the optimal time for that hardware configuration and for the stencil of
interest over all possible choices of tile sizes.

This results in the following optimization problem, which finds the best EH
parameters for a given stencil and a given bound on the chip area denoted by
$\mathit{chip\_area}$
\newcommand{\spc}{&& \!\!\!\!\!\!\!\!\!\!\!\!}
\begin{align}
& \underset{n_\mathrm{SM},n_\mathrm{V},M_\mathrm{SM},t_\mathrm{S_1},t_\mathrm{S_2},t_\mathrm{T}}{\text{minimize}}
& &\!\!\!\!\!T_\mathrm{alg}(n_\mathrm{SM},n_\mathrm{V},M_\mathrm{SM},t_\mathrm{S_1},t_\mathrm{S_2},t_\mathrm{T}) \label{eq:opt1}\\[2ex]
& \text{~~~~~~~subject to:}
\spc\mathcal{A}_\mathrm{tot} (n_\mathrm{SM},n_\mathrm{V},M_\mathrm{SM})\leq\mathit{chip\_area}\label{eq:opt2}\\
&\spc M_\mathrm{tile} \leq M_\mathrm{SM/threadblock}\label{eq:opt3}\\
&\spc k\leq \mathit{MTB}_\mathrm{SM}\label{eq:opt4}\\
&\spc k*M_\mathrm{tile} \leq M_\mathrm{SM}\label{eq:opt5}\\
&\spc k,t_\mathrm{S_1} \mbox{ -- integers}\label{eq:opt6}\\
&\spc n_\mathrm{V},t_\mathrm{S_2} \mbox{ -- multiple of 32}\label{eq:opt7}\\
&\spc M_\mathrm{SM} \mbox{ -- multiple of 48\,k}\label{eq:opt8}\\
&\spc n_\mathrm{SM},t_\mathrm{T} \mbox{ -- even}\label{eq:opt9}
\end{align}

Note that $T_\mathrm{alg}$ in \eqref{eq:opt1} is in fact a function not only
of the parameters $n_\mathrm{SM}$, $n_\mathrm{V}$, $M_\mathrm{SM}$,
$t_\mathrm{S_1}$, $t_\mathrm{S_2}$, and $t_\mathrm{T}$, but also of the other
parameters detailed elsewhere~\cite{prajapati2017simple}.  Also, for 3D stencils there is an additional parameter to function \eqref{eq:opt1} which is the tile size in the third space dimension $t_\mathrm{S_3}$.
We are only explicitly using those $T_\mathrm{alg}$ parameters that are needed
in the specific context.  Constraint \eqref{eq:opt2} guarantees that the total
area for the current hardware configuration, according to the area model, does
not exceed the optimization parameter \textit{chip\_area}.  Constraint
\eqref{eq:opt3} asks the memory required for each tile not to exceed
$M_\mathrm{SM/threadblock}$, the shared memory available for a threadblock,
while \eqref{eq:opt4} and \eqref{eq:opt5} restricts the number of tiles
simultaneously executed by an SM (using hyperthreading) to be no greater than
the max threadblocks an SM can handle and such that the total memory of all
such tiles is at most $M_\mathrm{SM}$. The values of $k$ and $t_\mathrm{S_1}$
can be only integers, and $n_\mathrm{V}$, $t_\mathrm{S_2}$, $M_\mathrm{SM}$,
$n_\mathrm{SM}$ and $t_\mathrm{T}$ should be multiples of 32, 32k, and 2,
respectively, by constraints \eqref{eq:opt5}--\eqref{eq:opt9}.  We require
$n_\mathrm{SM}$ to be even, $n_\mathrm{V}$ to be multiple of 32, and
$M_\mathrm{SM}$ to be multiple of 32k in order to be consistent with the
patterns for these parameters chosen by the manufacturers in existing
GPUs. For the tile sizes, we want $t_\mathrm{T}$ to be even as it is a
requirement for hybrid-hexagonal
tiling~\cite{grosser-etal-GPUhextile-CGO2014}, while $t_\mathrm{S_2}$ being a
multiple of 32 ensures that neighboring threads in ${S_2}$ dimension can fit
in a number of full warps, where each warp is a group of 32 threads.

The optimization problem \eqref{eq:opt1}--\eqref{eq:opt9} allows, given a
stencil of interest (say Jacobi 2D) and specific values of the problem size
parameters $S_i$ and $T$, to find a combination of software and hardware
parameters that results in optimal (smallest) run time. However, it is
unlikely that one may need a hardware that will be used for just a single
problem size. A more relevant problem is to optimize the hardware over a
set of problem sizes. Given a set of  sizes
$\Sz=\{(S_1^{(i)},S_2^{(i)},T^{(i)})\}$ and a frequency function $\fr$ such
that $\fr(\sz_i)$ is the expected frequency the size $\sz_i\in \Sz$ will be
encountered for the intended application,  the corresponding new objective becomes
\begin{equation}\label{eq:opt02}
\begin{aligned}
& \underset{n_\mathrm{SM},n_\mathrm{V},M_\mathrm{SM},\Ts}{\text{minimize}}
& \!\!\!\!&T_\mathrm{alg}^\mathrm{J}= 
\sum_{s\in \Sz}\fr(s)T_\mathrm{alg}(s),
\end{aligned}
\end{equation}
where $\Ts$ is the set of tile sizes, with each tile size given as a triple
$(t_\mathrm{S_1}, t_\mathrm{S_2},t_\mathrm{T})$,
and $T_\mathrm{alg}^\mathrm{J}$ ($J$ is for Jacobi) is the time-model
function, whose EH parameters are $n_\mathrm{SM}$, $n_\mathrm{V}$,
$M_\mathrm{SM}$, and whose ES parameters are $\Ts$. Hence, the optimization
problem \eqref{eq:opt02} has $|\Ts|+2=3|\Sz|+2$ decision variables.

In our experiments, we use as values for $S$ the set
$\Sz_S=\{4096,8192,12228,16384\}$ and for $T$ the set
$\Sz_T=\{1024,2048,4096,8192,16384\}$ and define
$\Sz=\{(S,T)~|~S\in \Sz_S, T\in \Sz_T, T\leq S\}$.  We use $T\leq S$ since it
is known from practice that no more than $S$ iterations are needed for
convergence.  One can check that $|\Sz|=16$ and hence the number of variables
for problem \eqref{eq:opt02} is $16*3+2=50$.  While this is a small number of
variables for optimization problems with a nice structure, e.g., linear or
convex, in our case the problem is of a difficult type, with all variables being
integer and the objective function and the constraints being rational
functions.  Because of the floor and ceiling functions used in our time model,
our objective and constraints are not even continuous, although this can be
handled by introducing one new integer variable per floor or ceiling function.
This however further increases the number of variables, which becomes
$16*10+2=162$.

Regardless of the fact that an optimization problem with 162 integer variables
and non-convex constraints and objective is in most cases computationally
infeasible, we will consider an even larger version
of \eqref{eq:opt02}, and then show how both problems can be simplified and
solved accurately and in reasonable time by exploring the separability of the objective.

Let us first describe a problem setting that is more relevant from practical
point of view.  While \eqref{eq:opt02} can
be used to determine \textit{stencil-optimal} architecture parameters, in many cases,
people would be interested in having a special-purpose hardware that is
optimized for an application whose computation time is dominated by a set of
often-used kernels.  In our case, we assume that a hypothetical application
$\mathit{Apl}$ is using the six stencils studied in this paper and by
studying a profiling data we have determined how often each stencil is used
and a distribution of problem dimension sizes for each application.  
In our experiments we use four 2D stencils:
Jacobi-2D, Heat-2D, Laplacian-2D, and Gradient-2D, all first order stencils, and two 3D stencils: Heat-3D and Laplacian-3D.  All 2D stencils
have two space dimensions and one time dimension, while the 3D stencils have three space dimensions and one time dimension.
We then
extend the definition of the frequency function so that, for each code
$c\in \Cd :=\{\mbox{Jacobi-2D, Heat-2D, Laplacian-2D, Gradient-2D, Heat-3D},$ $\mbox{Laplacian-3D}\}$, $\fr(c)$
denotes the frequency of using $c$ in application $\mathit{Apl}$.  Moreover,
given $c\in\Cd$ and $\sz\in\Sz$, let $\fr(c,\sz)$ denote the frequency with
which problem size $\sz$ has been used for stencil $c$. Then the updated
objective of the resulting optimization problem becomes
\begin{equation}\label{eq:opt03}
\begin{aligned}
& \underset{{\mathit{HP}},\Ts}{\text{minimize}}
& \!\!\!\!&T_\mathrm{alg}^\mathrm{\Cd}= 
\sum_{c\in\Cd,s\in \Sz }\fr(c)\fr(c,\sz)T_\mathrm{alg}^c({\mathit{HP}},\Ts,\sz).
\end{aligned}
\end{equation}

The set $\Ts$ here includes tile sizes for each $c\in \Cd$ and each
$\sz\in \Sz$. Hence the number of integer variables for problem \eqref{eq:opt03}
is $10|\Cd||\Sz|+2=10\cdot 4\cdot 16+2=642$, which is too large to be solved by existing solvers, given that the problem is nonlinear, nonconvex, and with integer variables.
Fortunately, the problem can be made feasible by dividing the variables into two sets and combining exhaustive search on the first set with nonlinear programming optimization on the second. 

\subsection{Solving the optimization problem}
The main observation that helps us solve problem \eqref{eq:opt03} more
efficiently is the fact that, if we knew the optimal value $\mathit{hp}_{opt}$ of parameters $\mathit{HP}$, then the objective \eqref{eq:opt03} would become separable since $T_\mathrm{alg}^c({\mathit{hp}_{opt}},\Ts,\sz)$ can be minimized with respect to the tile sizes independently for each $c$ and $\sz$. 
Formally, we are transforming \eqref{eq:opt03} into the following equivalent problem
\begin{equation}\label{eq:opt04}
\begin{aligned}
  & \underset{\mathit{hp}\in \mathit{HP}}{\text{minimize}} \sum_{c\in\Cd,\sz\in \Sz
  }\fr(c)\fr(c,s) ~~ \underset{\Ts (c, s)}{\text{min}}T_\mathrm{alg}^c({hp},\Ts,\sz).
\end{aligned}
\end{equation}

Since we don't know $\mathit{hp}_{opt}$, then we replace $\mathit{hp}$ in \eqref{eq:opt04} with all feasible values of $\mathit{HP}$ and, for each such value, we run a separate optimization problem with respect to $\Ts$ for each combination of $c\in \Cd$ and  $\sz\in \Sz$. 
As a result,
instead of solving one large problem \eqref{eq:opt03} with 642 variables, we
do exhaustive search on the space $\mathit{HP}\times \Cd\times \Sz$ and, for each point in that space, we solve using a nonlinear solver an optimization problem with only 10 integer variables.  (While the variables we are interested in are only three, the optimization problem includes additional internal variables such as the optimal number of tiles for hyperthreading and variables used to simplify some nonlinearities, resulting in total number of 10.) 


To further reduce the number of these problems, we will fix the range of values of the parameters as follows: $2 \leq n_\mathrm{SM} \leq 32$ and is even, $32 \leq n_\mathrm{V} \leq 2048 $ and is a multiple of 32, which come from the constraints of \eqref{eq:opt1}--\eqref{eq:opt9}.  $48k \leq M_\mathrm{SM} \leq 480k$ with the requirement \eqref{eq:opt8} to be multiple of 48k and positive.  In addition, we explore the sizes $12k$, $24k$ and $36k$ for $M_\mathrm{SM}$. 

\begin{figure*}[tb]
\centering
\includegraphics[width=3in]{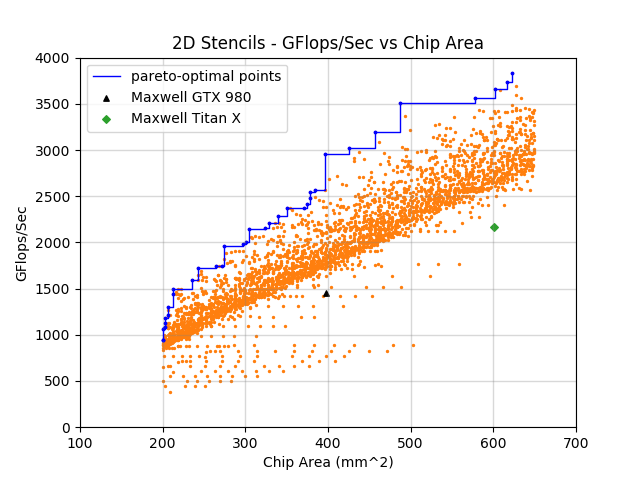}
\includegraphics[width=3in]{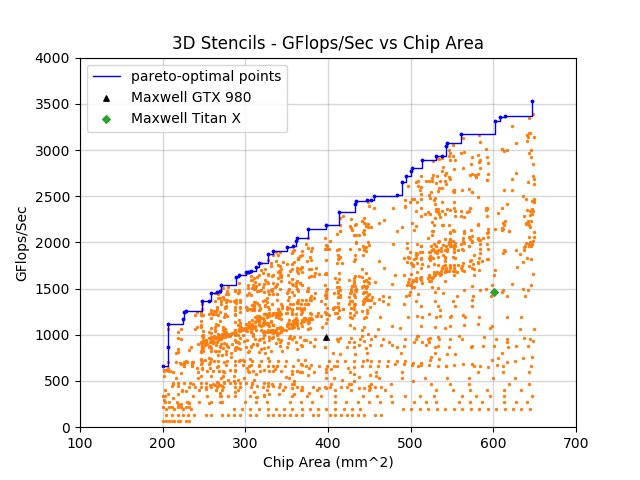}

\protect\caption{Optimal performance (in GFLOPs/Sec) of each feasible design
  point as a function of total chip area for 2D stencils (left) and 3D
  stencils (right).  Each benchmark and input size combination is assumed to
  occur with equal frequency.  Of the many thousands of points ($\approx 3000$
  for 2D stencils and $\approx 2000$ for 3D) many are \emph{dominated} by other design
  choices--a different architecture with smaller area yields better
  performance.  So, only a few tens of design points (the Pareto optimal
  points, shown in blue) need to be explored further, a nearly 100-fold savings
  in design cost.  In addition, the performance of the existing nVidia Maxwell
  GTX980 and Titan X are also shown.  The
  optimized designs for comparable area budget could improve performance of 2D stencils by as
  much as 104\% (c.f.\ 69\%) relative to the GTX980 (c.f., the TitanX). The performance of 3D stencils can be improved by as much as 123\% (c.f.\ 126\%) relative to the GTX980 (c.f., the TitanX).}
\label{fig:pareto-all}
\end{figure*}

We ran three experiments using chip areas in the range $200-650 \text{mm}^2$.
Since we haven't tied the experiments to a specific application, we assumed
all six stencils equally likely, and that each size combination also equally
likely; i.e., we set all coefficients in \eqref{eq:opt03} to 1.  
 For solving the optimization problems we used the open
source solver bonmin \cite{bonmin}.  The average solution time per
optimization instance is 19~sec (although it varies a lot from instance to
instance), so the total solution time varies between 7 and 24 hours depending
on the chip area.  Also note that, in the execution time model we use a
parameter $C_\mathrm{iter}$, the execution time of a single iteration on one
thread.  For optimal tile size selection, we measured this parameter for the
different stencils.  Although there was a small difference between the two
platforms (GTX-980 and Titan~X) we used the former value for our experiments.

\section{Discussion and Perspectives}
\label{sec:discussion}

We will now illustrate how our codesign optimization can be used to explore
various scenarios.  Furthermore, we will see that the trick of partitioning
the design space into a number of independent optimization problems, described
by Eqn.~(\ref{eq:opt04}), has added advantages of enabling investigation of a
number of scenarios without re-solving optimization problems.

\subsection{Design Space Exploration}

Given a range of area budgets between 200 and 650 $mm^{2}$, we enumerated all
feasible architecture, and solved the codesign optimization problem for each
one.  We separated the workload into two classes, 2D stencils and 3D stencils.
Each one illustrates interesting features.  For each class, we assumed a
uniform frequency (each instance is equally likely, and within each instance,
each problem size is also equally likely).  Figure~\ref{fig:pareto-all} shows
out results.  The first observation is that only about 1\% of the thousands of
feasible hardware designs (the Pareto optimal designs shown in blue) are worth
exploring further, leading to a significant pruning of the design space.

We also show in the same figure the performance of two standard design points,
Nvidia's GTX980 and TitanX architectures.  For comparable area, we can improve
performance by 104\% (resp., 69\% wrt.\ TitanX) for 2D stencils, and by 123\%
(resp., 126\%) for 3D stencils.  A large part of the gains come from the fact
that our proposed designs do not have caches because the HHC compiler for
which our model is specialized~\cite{grosser-etal-GPUhextile-CGO2014}
generates codes to perform data transfers explicitly rather than relying on
caches.  To compensate for this, we could ``delete'' the caches from the
GTX980 and TitanX, which reduces their areas to respectively, $237 \mathrm{mm}^2$ and
$356 \mathrm{mm}^2$.  The performance of the corresponding
Pareto optimal designs for those reduced areas is 9.34\% (resp., 28.44\% for the
cache-less TitanX) better for 2D stencils and 9.22\% (resp., 33.15\%) better for 3D
stencils.

\subsection{Workload Sensitivity}

Eqn.~(\ref{eq:opt04}) partitions the design space into a number of independent
optimization problems.  This allows us to explore other hypothetical scenarios
``for free.''  As an example, changing frequencies of the different programs
in the benchmark suite requires us to simply recompute new weighted sums of
optimal values of minimization problems that we have already solved.  A
specific example would be to hypothetically set the frequency for one of the
benchmarks as one (and zero for the others) thereby allowing us to explore
designs optimized for a single kernel.  It also helps determine whether the
chosen suite is representative of the mix that occurs in practice.
Table~\ref{configurations} illustrates the architectural parameters for
the best performing designs for each of the six benchmarks, for an area budget
between 425--450 $\mathrm{mm}^2$.

\begin{table}
  \centering
  \begin{tabular}{||l||r|r|r|r|r||}
    \hline\hline
    \multicolumn{1}{||c|}{Code} &
    \multicolumn{1}{|c}{$n_\mathrm{SM}$} &
    \multicolumn{1}{|c}{$n_\mathrm{V}$} &
    \multicolumn{1}{|c}{$M_\mathrm{SM}$} &
    \multicolumn{1}{|c|}{Area} &
    \multicolumn{1}{|c||}{GFLOPs/S} \\ \hline\hline
    Jacobi 2D & 32 & 128 & 24 & 438 & 2059 \\
    Heat 2D & 22 & 256 & 12 & 447 & 3017 \\
    Gradient 2D & 28 & 160 & 24 & 431 & 4963 \\
    Laplacian 2D & 28 & 160 & 12 & 426 & 2549 \\
    Heat 3D & 18 & 288 & 192 & 447 & 3600 \\
    Laplacian 3D & 8 & 896 & 96 & 446 & 1427 \\
    \hline\hline
  \end{tabular}
  \vspace*{2mm}
  \label{configurations}
  \caption{Workload sensitivity. The optimal architecture for a single
    benchmark is significantly different from that for others.}
  \label{tab:configurations}
\end{table}

Observe how the parameters of the best architecture are significantly
different.  There are also differences in the achieved performance for each
benchmark, but that is to be expected since the main computation in the
stencil loop body has different numbers of operations across the benchmark.
We can also observe that there is marked differences between 2D and 3D
stencils.  The latter seem to require larger shared memory, and well as higher
number of vector units.  Indeed, for designs will lower than 48kB, the
performance was nowhere near the optimal for these programs.

\begin{figure*}[tb]
\centering
\includegraphics[width=3in]{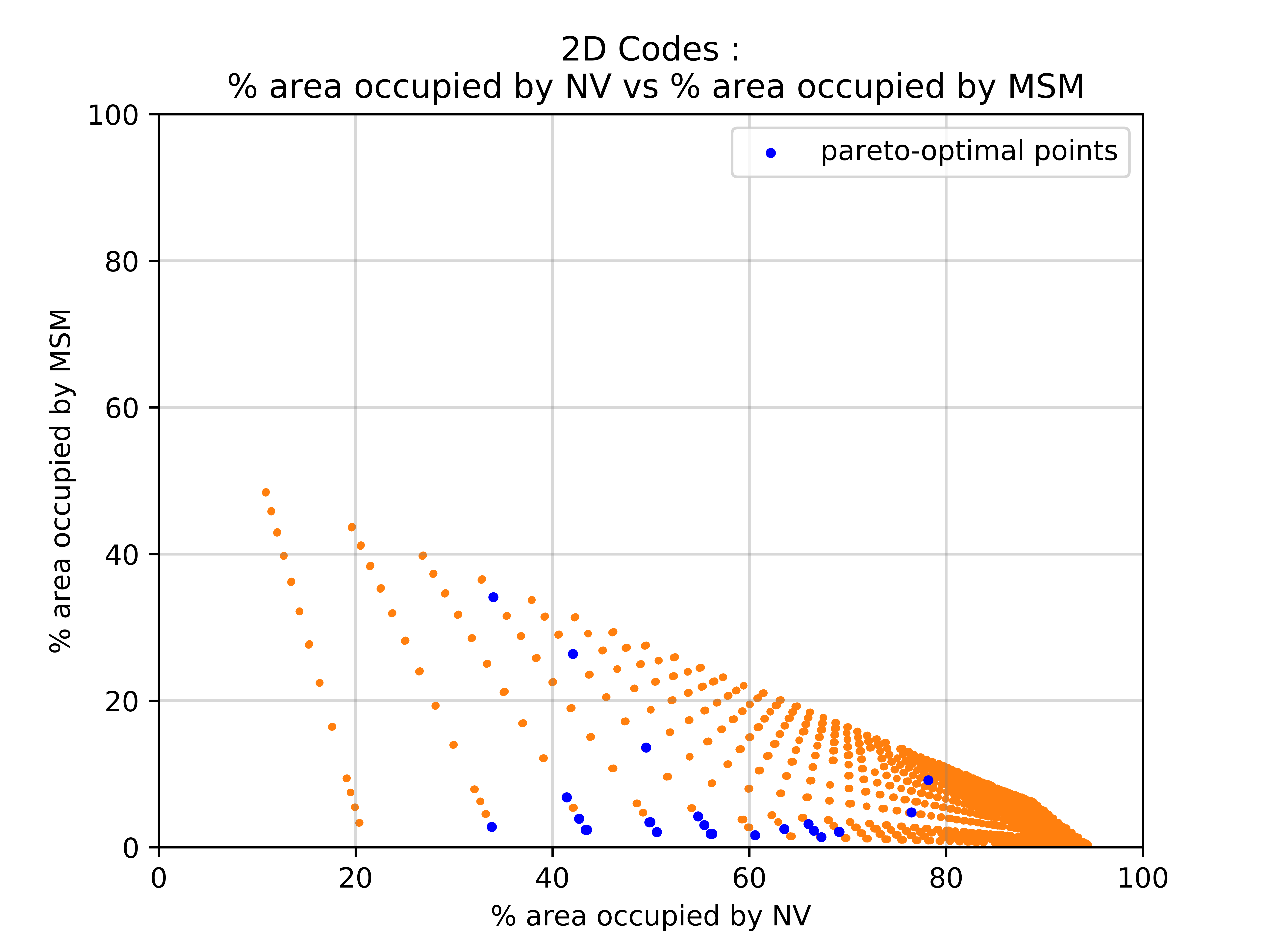}
\includegraphics[width=3in]{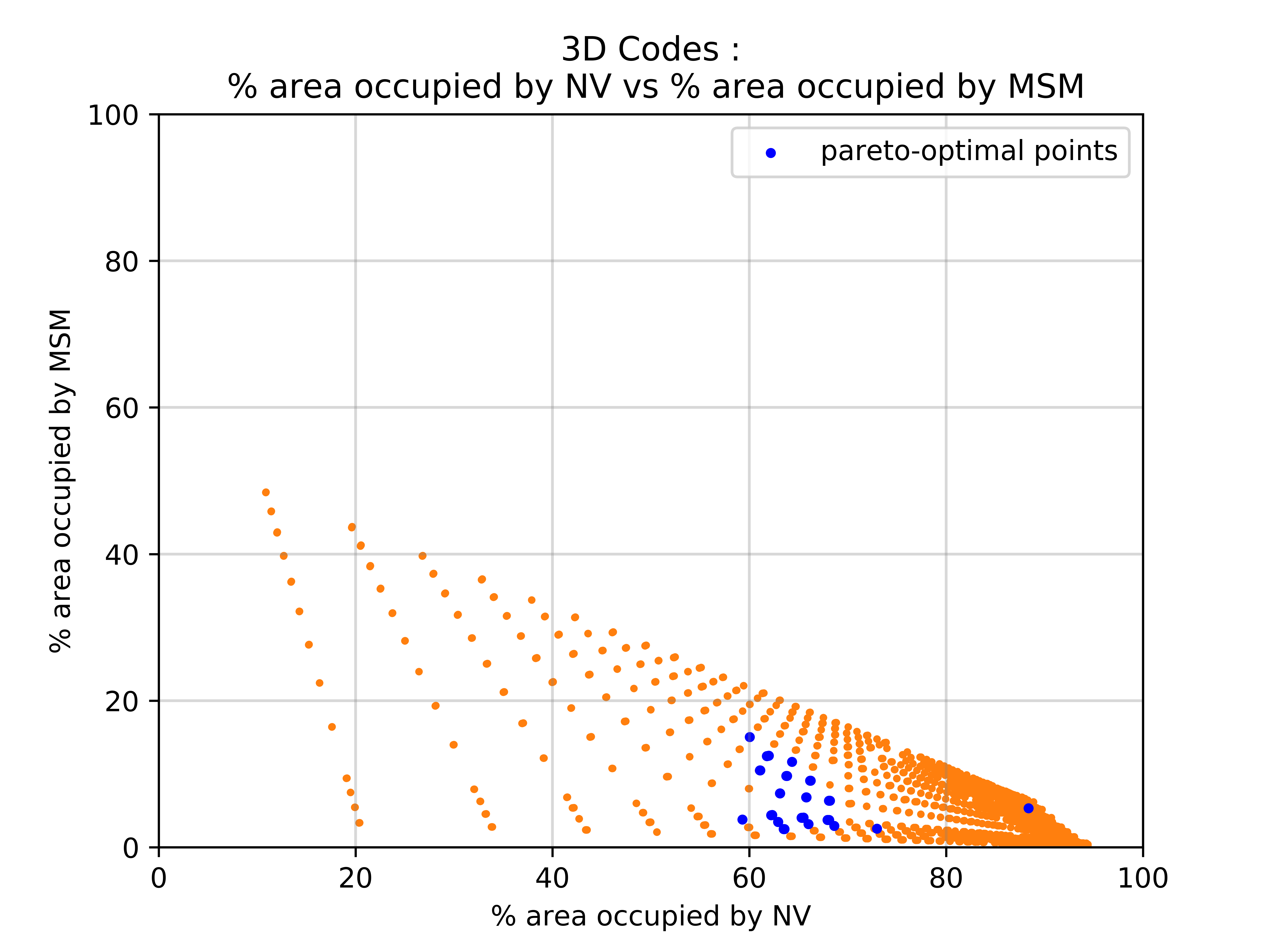}

\protect\caption{Resource Allocation.}
\label{fig:resource}
\end{figure*}

\subsection{Resource Allocation}
Another interesting perspective is seen in Figure~\ref{fig:resource}
which plots the same design space as in Figure~\ref{fig:pareto-all} but this
time the axes are different, showing the relative percentages of the chip area
devoted to memory, and to vector units.  We notice the that optimal designs
(blue points lie in a relative cluster.  This phenomenon is even more marked
for 3D stencils.  At present, we do not have a clear explanation for why the
points are clustered in this manner, an we plan to mine this data to determine
patterns in the data.

\subsection{Discussion}

The designer can choose to fix some parameters and optimize for others.  Given
an application, if the architecture parameters are fixed, the designer can use
our approach to optimize for compiler parameters.  On the other hand, if only
some of the architecture parameters are fixed, if the number of vector units
and the size of memories is fixed, then the designer can use our approach to
tune for the number of SMs.

One limitation of our current model is register usage. It is known to be
critical for performance, e.g., the size of the register file constrains the
optimal tile sizes. Currently, the register file size is a fixed constant in
the area model and does not appear in the execution time model. Modeling
these effects analytically is an ongoing effort.

One must, however, take this with a grain of salt.  Our models are approximate
and there are other parameters that can play an important role in the design
process.  For instance, the designer might be interested in optimizing for
both execution time as well as power simultaneously.  Our approach can be
extended to consider energy/power consumption into account.  If the energy
consumption details of the individual components of the architecture are
known, then the objective function can be updated to be the argmin of the
weighted execution times and energy components where we seek to maximize the
performance.  Such an optimization function can be formulated to solve
power-gating problems where the designer wants to turn off certain parts of
the chip.

Finally, we want to note two important aspects of this work.  First, our area
model is relatively simplistic, and second, the workload that we have chosen
is rather artificial.  Nevertheless, we have seen interesting patterns.  We
expect that designers in the field that have more precise (and probably
proprietary) information about their specific context could easily add
precision to our model and gather much more sophisticated conclusions.


\section{Related work}
\label{sec:related}
Our research draws upon prior work in three distinct areas.

\subsection{Chip Reverse Engineering and Area Modeling}

Chip area modeling can be formally considered a branch of semiconductor
reverse engineering, which is a well researched subject area.  Torrence
et.~al.~\cite{Torrence:2009} gives an overview of the various techniques used
for chip reverse engineering.  The packaged chips are usually decapped and the
wafer die within is photographed layer by layer.  The layers are exposed in
the reverse order after physical or chemical exfoliation. {\em
  Degate}~\cite{degate}, for example, is a well known open source software
that can help in analyzing die photographs layer by layer.  The reverse
engineering process can be coarse-grained to identify just the functional
macro-blocks.  Sometimes, the process can be very fine-grained, in order to
identify standard-cell interconnections, and hence, actual logic-gate
netlists.  Degate is often used in association with catalogs of known standard
cell gate layouts, such as those compiled by {\em Silicon
  Zoo}~\cite{SiliconZoo}.  Courbon et.~al.~\cite{Courbon:2016} provides a case
study of how a modern flash memory chip can be reverse engineered using
targeted scanning electron microscope imagery.  For chip area modeling, one is
only interested in the relatively easier task of demarcating the interesting
functional blocks within the die.

\subsection{Performance Modeling}

There have been various works on time modeling and performance optimization.

Polymage~\cite{Mullapudi2015Polymage} provides a stencil graph DSL and pairs it
it with a simple analytical performance model for
the automatic computation of optimal tile sizes and fusion choices.  With
MODESTO~\cite{Gysi2015Modesto} an analytical performance model has been
proposed that allows to model multiple cache levels and fusion strategies for
both GPUs and CPUs as they arise in the context of Stella.  For stencil GPU
code generation strategies that use redundant computations in combination with
ghost zones an analytical performance model has been
proposed~\cite{Meng2009GhostZhonePerfModel} that allows to automatically
derive ``optimal'' code generation parameters.  Yotov et.~al~\cite{Yotov2003}
showed already more than ten years ago that an analytical performance model
for matrix multiplication kernels allows to generate code that is
performance-wise competitive to empirically tuned code generated by
ATLAS~\cite{ClintWhaley20013}, but at this point no stencil computations have
been considered.  Shirako~et~al.~\cite{Shirako2012} use cache models to derive
lower and upper bounds on cache traffic, which they use to bound the search
space of empirical tile-size tuning.  Their work does not consider any GPU
specific properties, such as shared memory sizes and their impact on the
available parallelism.  In contrast to tools for tuning, Hong and
Kim~\cite{Hong2009} present a precise GPU performance model which shares many
of the GPU parameters we use.  It is highly accurate, low level, and requires
analyzing the PTX assembly code.  For stencil GPU code generation strategies
that use redundant computations in combination with ghost zones an analytical
performance model has been proposed~\cite{Meng2009GhostZhonePerfModel} that
allows to automatically derive ``optimal'' code generation parameters.

\subsection{Hardware/Software Codesign}


Application codesign is is a well established discipline and has seen active
research for well over two decades~\cite{Prakash-Parker1992, Wolf-1997,
  Chatha-Vemuri-Magellan-CODES2001, Dick-Jha-TCAD2006-mogac,
  Teich-proceedings2012}.  The essential idea is to start with a program (or a
program representation, say in the form of a CFDG---Control Data Flow Graph)
and then map it to an abstract hardware description, often represented as a
graph of operators and storage elements.  The challenge that makes codesign
significantly harder than compilation is that the hardware is not fixed, but
is also to be synthesized.  Most systems involve a search over a design space
of feasible solutions, and various techniques are used to solve this
optimization problem: tabu search and simulated
annealing~\cite{eles-etal-todaes97, Erbas-etal-TEC2006} integer linear
programming~\cite{Niemann-Marwedel1997}.

There is some recent work on accurately modeling the design space, especially
for regular, or \emph{affine control} programs~\cite{Pouchet-etalFPGA2013,
  Zuo-etal-ICCAD2013, Zuo-etal-CODES+ISSS2013}.  However, all current
approaches solve the optimization problem for a single program at a time.  To
the best of our knowledge, no one has previously considered the
\emph{generalized application codesign} problem, seeking a solution for a
\emph{suite of programs}.

There are multiple publications on codesign related to exascale computing, but
they focus on different aspects.  For instance, Dosanji et
a.~\cite{Dosanjh:2014:EDS:2562354.2562814} focus on methodological aspects of
exploring the design space, including architectural testbeds, choice of
mini-applications to represent applications codes, and tools.  The ExaSAT
framework \cite{doi:10.1177/1094342014568690} was developed to automatically
extract parameterized performance models from source code using compiler
analysis techniques.  Performance analysis techniques and tools targeting
exascale and codesign are discussed in
\cite{b1282de6ab8a4f1a8a962cb40310bc7f}.

\section{Conclusion}\label{sec:concl}
We develop a framework for software-hardware codesign that allows the simultaneous optimization of software and hardware parameters. It assumes having analytical models for performance, for which we use execution time, and cost, for which we choose the chip area. We make use of the execution time model from Prajapati et al~\cite{prajapati2017simple} that predicts the execution
times of a set of stencil programs. For the chip area, we develop an analytical model that estimates the chip area of parameterized designs from the Maxwell GPU architecture. Our model is reasonably accurate for estimating the total die area based on individual components such as the number of SMs, the number of vector units, the size of memories, etc.

We formulate a codesign optimization problem using the time model and our area model for optimizing the compiler and architecture parameters simultaneously.  We predict an improvement in the performance of 2D stencils by (104\% and 69\%) and 3D stencils by (123\% and 126\%) over existing Maxwell (GTX980 and Titan~X) architectures. 

The main focus of this paper is not on making specific design recommendations, but rather on the methodology; specifically, to develop a software-hardware codesign  framework and to illustrate how models built using it can be used for efficient exploration of the design space for identifying Pareto-optimal configurations and analyzing for design tradeoffs. The same framework, possibly with some modifications, could be used for codesign on other type of hardware platforms (instead of GPU), other type of software kernels (instead of the set of stencils we chose, or even non-stencil kernels), and other kind of performance and cost criteria (e.g., energy as cost). Also, with work focused on the individual elements of the framework, the execution time and the chip area models we used could possibly be replaced by ones with better features in certain aspects or scenarios. 

Despite this caveat, the analyses from this paper indicate the following accelerator design recommendations, for the chosen performance, cost criteria, and  application profile: 
\begin{itemize}
\item Remove caches completely and 
\item Use the area (previously devoted to caches) to add more cores on the
  chip.
\item The more precise the workload characterization and the specific area
  model parameters, the more useful the conclusions drawn from the study.
\end{itemize}


\bibliographystyle{plain}
\bibliography{references,codesign,bib4,parallelizing}

\end{document}